\documentclass{Interspeech2024}
\usepackage{multirow}
\usepackage{hyperref}
\usepackage{bbding}
\usepackage{booktabs}
\usepackage{booktabs}
\usepackage{cite}
\usepackage{amssymb}
\usepackage{threeparttable}
\usepackage{caption}

\interspeechcameraready

\title{SC-MoE: Switch Conformer Mixture of Experts for Unified Streaming and Non-streaming Code-Switching ASR}

\name{Shuaishuai Ye, Shunfei Chen, Xinhui Hu, and Xinkang Xu}

\address{HiThink RoyalFlush AI Research Institute, Zhejiang, China}
\email{\{yeshuaishuai, chenshunfei, huxinhui, xuxinkang\}@myhexin.com}

\keywords{automatic speech recognition, code-switching, streaming, mixture of experts, switch conformer}

\begin{document}
\maketitle
\begin{abstract}
In this work, we propose a Switch-Conformer-based MoE system named SC-MoE for unified streaming and non-streaming code-switching (CS) automatic speech recognition (ASR), where we design a streaming MoE layer consisting of three language experts, which correspond to Mandarin, English, and blank, respectively, and equipped with a language identification (LID) network with a Connectionist Temporal Classification (CTC) loss as a router in the encoder of SC-MoE to achieve a real-time streaming CS ASR system.
To further utilize the language information embedded in text, we also incorporate MoE layers into the decoder of SC-MoE. 
In addition, we introduce routers into every MoE layer of the encoder and the decoder and achieve better recognition performance. 
Experimental results show that the SC-MoE significantly improves CS ASR performances over baseline with comparable computational efficiency. 
\end{abstract}

\section{Introduction}
Code-switching ASR refers to the ability of a speech recognition system to recognize a conversation or utterance with alternation of two or more languages.
In a world where globalization and multilingualism are becoming increasingly prevalent, accurate and efficient CS ASR is critical. 
However, significant challenges remain in CS ASR due to 
the phonemic confusion among multiple languages\cite{liuhexin2023,boundary_asru2023,boundary_fanzhiyuan_Interspeech2023}. 
To address the data scarcity issues, various data augmentation approaches have been proposed, such as 
generating CS audio data\cite{audiosplicing2021,speechedit2023} and CS text data\cite{machineTranslate2021,yuhaibin2023,LLM}.
Additionally, self-supervised fine-tuning methods are also proposed \cite{selfsupervise_2023_1,selfsupervise_2023_2}.
To solve the phonemic confusion issues, the Mixture of Experts (MoE) has become the mainstream model architecture for CS ASR system due to its ability not only to make use of contextual information between different languages, but also to discriminate different language information and extract language-specific representation\cite{lae_2022,LAESTMoE,sw_wangyi_lr_moe_2023,sw_pingan_2023}. 

MoE models for CS ASR can be mainly divided into bi-encoder-based MoE
\cite{biencoder_luyizhou_2020} and Switch-Transformer-based MoE\cite{SwitchTransformer}.
The bi-encoder-based MoE is structured by decomposing the single transformer encoder originally used for monolingual ASR tasks into two separate language-specific encoders. 
It has been shown to have advantages over single-encoder models for CS ASRs
\cite{biencoder_2019,biencoder_2020,biencoder_song22e_Interspeech,lae_2022,LAESTMoE} . 
In order to address the limitations of bi-encoder-based MoE models, which often suffer from insufficient interaction between the separate encoders and neglect linguistic general representation, a language-aware encoder (LAE) \cite{lae_2022,LAESTMoE} introduces a shared encoder block before the top-level monolingual encoders. This shared block efficiently captures general representations that are common across different languages. Additionally, to leverage the full potential of contextual information between different languages, \cite{LAESTMoE} incorporates a speech translation task into LAE, enabling it to learn this information more effectively.

Switch-Transformer-based MoE is another novel MoE model architecture,
which not only mitigates the additional computational cost issue appeared in bi-encoder-based MoE, but also achieves better recognition performance\cite{sw_wangyi_lr_moe_2023,sw_pingan_2023,sw_mole_2023,google_conformer_moe2023}. 
However, the prior works still have some drawbacks, either ignoring language information or failing to achieve a real-time streaming ASR system. 
%
Studies like \cite{sw_pingan_2023,sw_mole_2023,google_conformer_moe2023} demonstrate real-time streaming capabilities. However, incorporating language information could further enhance their accuracy.
%
One approach, exemplified by \cite{sw_wangyi_lr_moe_2023}, utilizes a Language Identification (LID) network with CTC loss to achieve good recognition performance. However, its reliance on a simplified, rule-based alignment strategy for handling blank tokens hinders real-time streaming functionality.
%
The development of a robust ASR system that excels in both high accuracy and real-time streaming capabilities remains crucial. Such a system would unlock significant value in diverse applications, including human-computer interaction, real-time subtitles, and seamless meeting transcriptions.

In this paper, we propose a Switch-Conformer-based MoE (SC-MoE) system that aims to achieve high recognition accuracy with the support of language information while also enabling a streaming CS ASR system that allows users to control the latency at inference time.
To achieve a real-time streaming CS ASR system when employing LID-CTC loss, we consider blank as a language and design a streaming MoE layer composed of three language experts: Mandarin, English and blank, and a router. 
To leverage language-specific information from the text, we also incorporate MoE layers into the decoder of SC-MoE. 
Additionally, as verified in the experimental section, we introduce routers into every MoE layer of the encoder and the decoder.
Experimental results show that the proposed SC-MoE significantly improves CS ASR performances over baseline with comparable computational costs.

The contributions of this work are summarized as follows: 
(1) We propose a SC-MoE model for unified streaming and non-streaming CS ASR, 
which significantly outperforms the baseline systems with absolute Mixed Error Rate (MER) reductions of 0.98\% and 0.97\% in streaming and non-streaming modes, respectively.
(2) We design a streaming MoE layer composed of three language experts and a dedicated router with a LID-CTC loss to achieve a real-time streaming CS ASR system. 
(3) We introduced routers into every switch conformer encoder layer and switch transformer decoder layer, resulting in good recognition performances. 
\section{Related work}
\subsection{Switch Transformer}
Switch Transformer (ST) \cite{SwitchTransformer} is a sparsely-activated transformer model with a large number of parameters but a constant computational cost. 
Each encoder layer of ST is structured by replacing the dense feed-forward network (FFN) layer with a sparse switch FFN (sFFN) layer, resulting in two parts: shared parts and specialized parts.
A sFFN layer comprises $n$ expert layer(s) and a router. 
The sparsity of ST comes from activating the top-1 expert of sFFN layer with a maximum router gate value for each sample. 
The sFFN layer returns top-1 expert's output multiplied by the maximum router gate value. 
Additionally, ST incorporates an auxiliary balanced loss to encourage a balanced load across experts. 
For each sFNN layer, this auxiliary loss is added to the total model loss during training.

In CS and multilingual ASR, both the utilization of the contextual information across different languages and the extraction of the language-specific representations contribute to achieve good recognition performance\cite{sw_wangyi_lr_moe_2023,sw_mole_2023}.
Compared with the bi-encoder-based MoE with a shared encoder block, ST not only has the ability to boost the contextual information interaction via shared parts and extract language-specific representations through specialized parts but also requires lower computational costs \cite{sw_wangyi_lr_moe_2023}.
So, LR-MoE\cite{sw_wangyi_lr_moe_2023} adopted ST as the foundation of CS and multilingual ASR, and significantly improved recognition performance over standard Transformer models
with comparable computational complexity.

\subsection{Language-Routing Mixture of Experts: LR-MoE}
The architecture of LR-MoE \cite{sw_wangyi_lr_moe_2023} is structured by replacing the balanced loss in Switch Transformer with LID-CTC loss that solves the problem of difficulty in obtaining frame-level language labels.
In LR-MoE, the router regarded as a LID network was jointly trained with the ASR model using language labels obtained by replacing the tokens in the text labels with the corresponding
language.
To ensure that all routing paths of MoE layers in encoder are the same when inputting a sample, the LR-MoE adopts the strategy of a shared LID router, which can reduce multi-level error accumulation. 
While LR-MoE based on Switch Transformer has acquired great achievements, there exists some limitations as follows:
\begin{itemize}
    \item Blank replacement rule prevents LR-MoE from performing real-time streaming ASR, due to the uncertain response time of the first packet. Furthermore, one sample cannot be decoded when all outputs of LID are blank under such a rule in more extreme situations \cite{sw_wangyi_lr_moe_2023}.
    \item While the strategy of shared LID router can reduce multi-level error accumulation, it loses the opportunity for the following routers to correct errors, which results in "one mistake, all mistakes".
    \item The decoder lacks the modeling ability for phonemic confusion when utilizing only MoE layer in the encoder.
\end{itemize}
In order to address the above limitations, we propose SC-MoE for unified streaming and non-streaming CS ASR. 

\begin{figure}[t]
  \captionsetup{font=footnotesize}
  \centering
  \includegraphics[width=\linewidth]{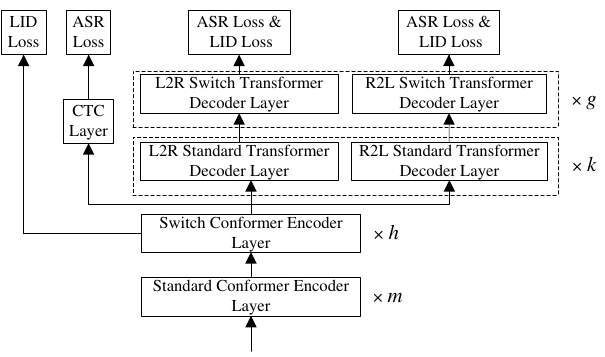}
    \vspace{-0.7cm}
  \caption{The architecture of the proposed SC-MoE. m, h, k, g represent the number of different network layers.}
  \label{fig:sc-moe}
    \vspace{-0.61cm}
\end{figure}

\section{SC-MoE: Our Proposed Model}

Figure.\ref{fig:sc-moe} shows an overview of the architecture of the proposed SC-MoE. 
We employ the U2++\cite{U2++}, a two-pass CTC and Attention Encoder Decoder (AED) joint architecture\cite{ctc_attention} as our backbone.
In order to improve the modeling abilities for phonemic confusion, we not only incorporate the MoE layer into the encoder, but also the decoder of the SC-MoE, which is different from LR-MoE. 
Both the encoder and decoder are structured by replacing the standard conformer layer\cite{conformer} and the standard transformer decoder layer with Switch Conformer (SC) encoder layer and ST decoder layer, respectively.

\subsection{Streaming MoE layer}
Due to the sparse spike property of the CTC\cite{CTC}, the output of the LID network (router) with CTC loss will contain a large amount of blank symbols. 
In LR-MoE, the authors utilize a Blank Replacement Rule (BRR) (detailed in \cite{sw_wangyi_lr_moe_2023}) to acquire dense frame-wise language routing information. However, this approach suffers from two key drawbacks:
(1) Uncertain first packet response: The BRR can lead to an unknown response time for the first packet, hindering the implementation of a streaming CS ASR system. 
(2) Decodability issues: In extreme scenarios, all LID outputs might be blanks, rendering the sample undecodable under the given rule. 

To achieve a real-time streaming CS ASR system with LID-CTC loss, we propose abandoning the BRR \cite{sw_wangyi_lr_moe_2023} and instead, consider blank as a separate language. 
We introduce a novel streaming MoE layer (Figure \ref{fig:enc_dec} (a)), composed of three language experts: Mandarin, English and blank, and a dedicated router. 
A linear router selects and activates the expert with the highest output probability to extract language-specific representations efficiently. 
This approach eliminates the need for the problematic BRR, enabling a truly streaming ASR system. 
Notably, compared to the Mixture of Language Expert (MLE) FNN layer in \cite{sw_wangyi_lr_moe_2023}, our streaming MoE layer introduces an additional blank expert, addressing the limitations of the original rule.

Given that the output of the $l$-1-th encoder layer $O^{l-1}$, the output of the previous layer of a streaming MoE layer is $O_p^l$, the $i$-th expert weights, the router weights and the router biases of a streaming MoE layer in the $l$-th switch conformer encoder block are $W_{i}^l$, $W_{r}^l$ and $b_r^l$, respectively. Then, the output $O_{s}^l$ of the streaming MoE layer can be formulated as:

\begin{equation}
    O_{s}^l={p_i^l}({O_p^l*W_{i}^l})
\end{equation}
\begin{equation}
    i=argmax(p_i^l)
\end{equation}
\begin{equation}
    p_i^l=\frac{e^{o_{ri}^l}}{{\sum_{j=1}^3}e^{o_{rj}^l}}
\end{equation}
\begin{equation}
    O_r^l=O^{l-1}*W_r^l+b_r^l
\end{equation}
Where $O_r^l$ is the logits produced by the router and are normalized via a softmax distribution over the available three experts. $p_i^l$ is the gate-value for the expert $i$.

\subsection{Switch Conformer Encoder}
%
As depicted in Figure \ref{fig:enc_dec} (a), an SC encoder layer is modified by replacing two dense FFN layers in a standard conformer\cite{conformer} layer with two sparse streaming MoE layers (represented by the red dashed box).
Unlike \cite{sw_wangyi_lr_moe_2023}, we incorporate LID networks equipped with CTC losses into each SC encoder layer. 
This approach mitigates the "one mistake, all mistakes" issue prevalent in LR-MoE by offering subsequent routers the ability to rectify errors. 
To guarantee identical routing paths for the two streaming MoE layers within the same encoder layer, a single LID network (router) is shared between them. 
Each MoE layer processes the output of the preceding encoder layer as its input.

\subsection{Switch Transformer Decoder}
\begin{figure}[t]
  \captionsetup{font=footnotesize}
  \centering
  \includegraphics[width=\linewidth]{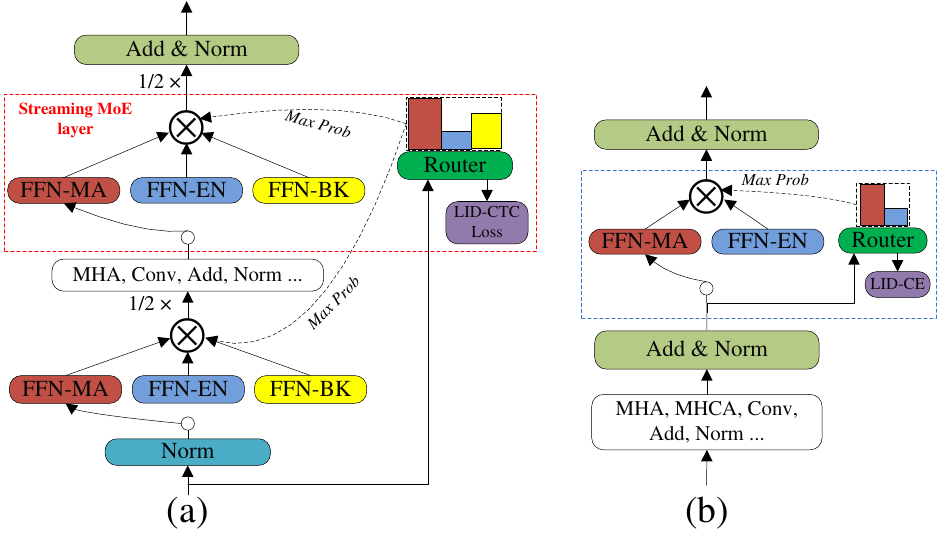}
    \vspace{-0.6cm}
  \caption{ (a) A Switch Conformer encoder layer of the proposed SC-MoE. (b) A Switch Transformer decoder layer of the SC-MoE. MA, EN and BK represent Mandarin, English and blank respectively. For simplicity, all skipping connection lines have been omitted.}
  \label{fig:enc_dec}
    \vspace{-0.4cm}
\end{figure}

In the ST decoder layer of SC-MoE, we adopt the Switch Transformer architecture, as illustrated in Figure.\ref{fig:enc_dec} (b). Instead of the original balanced loss, we employ cross-entropy (CE) loss specifically tailored for the 2-class LID task (distinguishing Mandarin and English). This integration of MoE layers (represented by the blue dashed box) within the decoder enhances the CS ASR model's ability to differentiate phonemes across languages.
\begin{table*}[th]
\renewcommand{\arraystretch}{0.96}
\captionsetup{font=footnotesize}
\caption{The performance (WER/CER/MER(\%)) of our proposed method and baselines reimplemented using WeNet toolkit. 'Man' is the mono-Mandarin CER, 'Eng' is the mono-English WER, and MER is the code-switching MER. 'INIT.' means using NativeConformer model to initialize the encoder and/or decoder layer of a model. 'SC-MoE-Enc' and 'SC-MoE-Enc-Dec' represent only employing MoE layer to the encoder and to both the encoder and decoder, respectively. 'R1', 'R2' and 'R3' represent three types of router, and their specific meanings can be found in the experimental section. [x, y] represents that chunk size is x, and the number of left chunk is y.}
\vspace{-0.2cm}
\label{table1}
\centering
\begin{tabular}{c|c|c|c|c|c|ccc|ccc}
\toprule
 \multirow{2.5}{*}{No.} & \multirow{2.5}{*}{Model}    &\multirow{1.5}{*}{Router}   &\multirow{2.5}{*}{INIT.} &\multirow{1.5}{*}{Model}       &\multirow{1.5}{*}{Activated}    &\multicolumn{3}{c|}{Streaming [16, 8]}       &\multicolumn{3}{c}{Non-streaming [-1, -1]}  \\ \cmidrule{7-12}
       
      &             &\multirow{-1.5}{*}{Type}   &                        & \multirow{-1.5}{*}{Parameters (MP)} &\multirow{-1.5}{*}{MP}        & Man      & Eng       & Mixed     & Man      & Eng       & Mixed     \\ 
\midrule
 1 & NativeConformer                   & &\tiny{\XSolidBrush}         &  50.2M                 &  50.2M                 & 7.95     & 30.53     & 10.40     & 7.33   & 28.55    & 9.63    \\
\midrule
2 & LAE                       & & \tiny{\XSolidBrush}         &  51.6M                 &  50.2M                 & 7.44     & 30.60     & 9.89      & 6.84   & 28.22    & 9.16    \\ 
\midrule
   3 &       LR-MoE*                & R1 & {\checkmark}           &     62.8M          &      50.2M       & 7.34     & 30.52     & 9.85      & 6.72   & 28.51    & 9.08    \\
\midrule
4 & \multirow{4}{*}{SC-MoE-Enc}  & R1 &{\checkmark}   & \multirow{4}{*}{75.4M}      & \multirow{4}{*}{50.2M}   & \textbf{7.12}    & 29.83     & 9.59      & \textbf{6.47}   & 27.96    & 8.80 \\
5 &                            & R2 &{\checkmark}           &                  &                    & 7.36     & 29.01     & 9.71      & 6.85   & 27.64    & 9.10 \\
6 &                           & R3 &\tiny{\XSolidBrush}           &                  &                     & 7.38     & 29.17     & 9.74      & 6.79   & 27.49    & 9.03 \\
7 &                           & R3 &{\checkmark}           &                  &                     & 7.16     & 28.68     & 9.50      & 6.48   & 26.95    & 8.70 \\

\midrule
 8 &        SC-MoE-Enc-Dec                 & R3 &{\checkmark}           &      83.8M                  &      50.2M                  & 7.15     & \textbf{28.05}     &\textbf{9.42}     &6.50   & \textbf{26.36}    & \textbf{8.66}    \\

\bottomrule
\end{tabular}
\begin{tablenotes}
\footnotesize
\item[*] {* Pseudo streaming without considering the response time of the first packet.}
\vspace{-0.5cm}
\end{tablenotes}
\end{table*}

\subsection{Training Objectives}
The training process is to jointly optimize the weight-sum of the CS ASR loss $\mathcal{L}_{asr}$ and the LID loss $\mathcal{L}_{lid}$. 
Given the acoustic features $x$, the corresponding annotations $y$, the language labels $z$ 
, the input of the LID (router) in the $i$-th SC encoder layer and the $j$-th ST decoder layer $h_{re}^i$ and $h_{rd}^j$ respectively. The full loss $\mathcal{L}$, the $\mathcal{L}_{asr}$ and the $\mathcal{L}_{lid}$ can be formulated as:
\begin{equation}
    \mathcal{L}=\mathcal{L}_{asr}(x,y)+\mathcal{L}_{lid}(x,z)
\end{equation}
\begin{equation}
    \mathcal{L}_{asr}=\lambda\mathcal{L}_{asr\_ctc}(x, y)+(1-\lambda)\mathcal{L}_{asr\_ce}(x, y)
\end{equation}
\begin{equation}
    \mathcal{L}_{lid}=\lambda\sum_{i=m+1}^{m+h}\mathcal{L}_{lid\_ctc}^i(h_{re}^i,z)+(1-\lambda)\sum_{j=k+1}^{k+g}\mathcal{L}_{lid\_ce}^j(h_{rd}^j,z)
\end{equation}
\begin{equation}
    \mathcal{L}_{asr\_ce}=(1-\alpha)\mathcal{L}_{l2r\_asr}(x,y)+\alpha\mathcal{L}_{r2l\_asr}(x,y)
\end{equation}
\begin{equation}
    \mathcal{L}_{lidce}=(1-\alpha)\mathcal{L}_{l2r\_lid}(h_{rd},z)+\alpha\mathcal{L}_{r2l\_lid}(h_{rd},z)
\end{equation}
Where $\mathcal{L}_{asr\_ctc}$ is the ASR CTC loss, $\mathcal{L}_{asr\_ce}$ is the ASR CE loss of the AED model, $\mathcal{L}_{lid\_ctc}^i$ is the LID CTC loss of $i$-th SC encoder, $\mathcal{L}_{lid\_ce}^j$ is the LID CE of $j$-th ST decoder. $\mathcal{L}_{l2r\_asr}$ and $\mathcal{L}_{r2l\_asr}$ are the ASR CE losses of the L2R and the R2L AED model respectively. $\mathcal{L}_{l2r\_lid}$ and $\mathcal{L}_{r2l\_lid}$ are the LID CE losses of the L2R and the R2L AED model respectively. 
$\lambda$ and $\alpha$ are hyper-parameters used to balance different part of losses.

\section{Experiments}
\subsection{Datasets} 
The CS ASR experiments are conducted on the 1160-hour Mandarin-English dataset. This dataset is composed by combining the ASRU 2019 Mandarin-English Code-Switching dataset (ASRU2019 for short)\cite{ASRU} and the train\_clean dataset of LibriSpeech dataset\cite{2015Librispeech}. 
ASRU2019 contains 200-hour code-switching speech and 500-hour Mandarin speech and the train\_clean dataset of LibriSpeech contains 460-hour mono-English data.
The test set for the experiments comprises a 20-hour code-switching speech data set from the ASRU2019.

\subsection{Experimental Setups}
We conducted all experiments on the WeNet\cite{wenet} toolkit and adopted the U2++\cite{U2++} as our backbone. 
We used the 80-dimensional log-Mel filter-bank feature as acoustic features for input, and these features were extracted from every speech frame with a frame-length of 25ms and a frame-shift of 10ms. The cepstral mean and variance normalization and SpecAugment\cite{specaugment} are applied in our all experiments.

Our proposed SC-MoE is configured with a 12-layer encoder, a 3-layer left-to-right (L2R) decoder, a 3-layer right-to-left (R2L) decoder and a CTC layer. 
The 12-layer encoder comprise a 6-layer standard conformer encoder and a 6-layer switch conformer encoder. 
Both L2R and R2L decoders comprise a 1-layer standard transformer decoder and a 2-layer switch transformer decoder. 
To demonstrate the effectiveness of the SC-MoE, we select three baseline systems with a similar computational cost for a fair comparison: 
(1) NativeConformer\cite{U2++}: 12 stacked conformer encoder layers, 
(2) Language-Aware Encoder (LAE)\cite{lae_2022}: The shared block contains 6 conformer layers , and the language-specific blocks consist of 3 conformer layers each, and 
(3) LR-MoE\cite{sw_wangyi_lr_moe_2023}: The LR-MoE encoder consists of a 6-layer standard conformer encoder and a 6-layer MLE conformer encoder, where both two FNN layers are all replaced with MLE FNN layer.  
All baseline systems adopted U2++ model with a 3-layer L2R decoder and a 3-layer R2L decoder. 
According to \cite{sw_wangyi_lr_moe_2023,wang21z_interspeech}, we adopt the first 6-layer conformer encoder layers and/or the first transformer decoder layer of NativeConformer model as pre-trained model to initialize the corresponding layers of the SC-MoE and LR-MoE.

All models are trained using Adam optimizer for 120 epochs with warm-up steps of 25k and peak learning rate of 0.001 on 4 GeForce RTX 3090 GPUs. We adopt the training scheme of dynamic chunk and dynamic left chunk with a dynamic batch size of a maximum frames size of 38000 (380s) to train models. In the process of optimizing model parameters, the loss-balanced hyper-parameters are set as follows: $\lambda$=0.3, $\alpha$=0.3 in training stage, while $\lambda$=0.3, $\alpha$=0.6 in decoding stage.  We averaged the best 10 models based on the loss value as the final model for decoding, and we adopted the decoding strategy of 'attention\_rescoring' in WeNet for all our experiments. 
\subsection{Experimental Results}
\subsubsection{Comparisons between SC-MoE and Baselines}
Table \ref{table1} compares the performance of our proposed SC-MoE against three baseline systems under the comparable computational costs. 
For a fair comparison with LR-MoE in streaming mode, we excluded the response time of the first packet.
The proposed SC-MoE-Dec-Enc (No.8) demonstrates superior performance compared to three baselines, particularly in mono-English test set, where it achieves an absolute Word Error Rate (WER) reduction of about 2.5\%. 
Notably, compared to the Switch-Transformer-based LR-MoE (No.3), the SC-MoE-Dec-Enc (No.8) also achieves a relatively significant improvement in both streaming and non-streaming modes.
In streaming mode, our method surpasses NativeConformer by achieving an absolute Character Error Rate (CER) reduction of 0.8\% (from 7.95\% to 7.15\%) for the mono-Mandarin test set, WER reduction of 2.48\% (from 30.53\% to 28.05\%) for the mono-English test set, and Mixed Error Rate (MER) reduction of 0.98\% (from 10.40\% to 9.42\%) for the CS test set.
In non-streaming mode, our method achieves CER/WER/MER of 6.50\%/26.36\%/8.66\%,
which showed to be superior to other models.

\subsubsection{Ablation Studies}
%
In addition to comparing the performance of our proposed SC-MoE with baseline systems in Table~\ref{table1}, we further investigated the effectiveness of the streaming MoE layer and the impact of:
Initialization using a pre-trained model, 
Different router types, and 
MoE layer incorporation in the decoder.

\textbf{Streaming MoE layer}: 
Compared to LR-MoE (No.3), SC-MoE-Enc (No.4) consistently surpasses its performance in both streaming and non-streaming modes. 
This suggests that the streaming MoE layer, equipped with an additional blank expert, outperforms LR-MoE's blank replacement rule, facilitating a real-time streaming system.

\textbf{Initialization}: 
Initializing SC-MoE-Enc with a pre-trained model (Rows 6 and 7) leads to an absolute MER reduction of 0.14\% and 0.33\% for streaming and non-streaming modes, respectively. 
This suggests that initialization improves the initial classification performance of the router, indirectly enhancing the CS ASR model's recognition performance. 
Notably, even without initialization (No.6), our model outperforms LR-MoE (No.3). Therefore, to achieve comparable recognition performance, our model can eliminate the need for pre-trained models, consequently reducing GPU requirements and shortening training time.

\textbf{Router type}: 
This section investigates the impact of three router configurations on the CS ASR model's performance:
%
%
(1) R1: All streaming MoE (sMoE) layers in SC-MoE or all MLE FNN layers in LR-MoE share the same router.
(2) R2: Each sMoE layer in SC-MoE has its own independent router.
(3) R3: Two sMoE layers from the same SC encoder layer share the same router.
Compared to configurations No.4 and No.5, No.7 with R3 consistently achieves the best performance in both streaming and non-streaming modes. 
Compared to No.4 with R1, it is suggested that No.7 with R3 mitigates the "one mistake, all mistakes" issue observed in LR-MoE. 

On the contrary, the performance of the R1 is better than the R3 in LR-MoE\cite{sw_wangyi_lr_moe_2023}.
The blank replacement rule used in LR-MoE likely results in multi-level error propagation, which stems from errors introduced not only by the router itself but also by the inherent limitations of the rule.
In contrast, our SC-MoE (No.7 with R3) avoids error propagation beyond the language identification (LID) module, eliminating the possibility of additional errors introduced by a rule-based approach. Compared to No.5 with R2, No.7 with R3 achieves an absolute MER reduction of 0.21\% (streaming) and 0.4\% (non-streaming).
R2's slightly lower performance is likely attributable to the two sMoE layers within the same SC encoder layer potentially taking different routing paths for a single sample.

\textbf{Decoder with a MoE layer}: 
Compared to SC-MoE-Enc (No.7), SC-MoE-Enc-Dec (No.8) achieves modest MER reductions in both streaming and non-streaming modes, albeit with a slight CER increase for mono-Mandarin test set in streaming mode. 
The results show that introducing MoE layers into the decoder potentially allows for utilizing language-specific text information, mitigating some phonemic confusion issues.

\section{Conclusions}
This paper introduces a novel Switch-Conformer-based Mixture of Experts (SC-MoE) model designed for unified streaming and non-streaming Continuous Speech Recognition (CSR) tasks.
%
To achieve real-time streaming in a LID-CTC loss setting, we propose a streaming MoE layer. As verified in the experimental section, this layer consists of three language experts (Mandarin, English and blank) and a dedicated router.
Experimental results demonstrate that SC-MoE significantly outperforms baseline systems with comparable computational efficiency in both streaming and non-streaming modes. 
Notably, in non-streaming mode, our method achieves CER/WER/MER of 6.50\%/26.36\%/8.66\%. In the future, we will explore the effectiveness of our proposed SC-MoE on multilingual ASR.
\bibliographystyle{IEEEtran}
\bibliography{mybib}

\begin{thebibliography}{10}
\providecommand{\url}[1]{#1}
\csname url@samestyle\endcsname
\providecommand{\newblock}{\relax}
\providecommand{\bibinfo}[2]{#2}
\providecommand{\BIBentrySTDinterwordspacing}{\spaceskip=0pt\relax}
\providecommand{\BIBentryALTinterwordstretchfactor}{4}
\providecommand{\BIBentryALTinterwordspacing}{\spaceskip=\fontdimen2\font plus
\BIBentryALTinterwordstretchfactor\fontdimen3\font minus \fontdimen4\font\relax}
\providecommand{\BIBforeignlanguage}[2]{{%
\expandafter\ifx\csname l@#1\endcsname\relax
\typeout{** WARNING: IEEEtran.bst: No hyphenation pattern has been}%
\typeout{** loaded for the language `#1'. Using the pattern for}%
\typeout{** the default language instead.}%
\else
\language=\csname l@#1\endcsname
\fi
#2}}
\providecommand{\BIBdecl}{\relax}
\BIBdecl

\bibitem{liuhexin2023}
H.~Liu, H.~Xu, L.~P. Garcia, A.~W.~H. Khong, Y.~He, and S.~Khudanpur, ``Reducing language confusion for code-switching speech recognition with token-level language diarization,'' in \emph{ICASSP}, 2023, pp. 1--5.

\bibitem{boundary_asru2023}
P.~Chen, F.~Yu, Y.~Liang, and et~al., ``Ba-moe: Boundary-aware mixture-of-experts adapter for code-switching speech recognition,'' \emph{Automatic Speech Recognition and Understanding (ASRU)}, 2023.

\bibitem{boundary_fanzhiyuan_Interspeech2023}
Z.~Fan, L.~Dong, C.~Shen, Z.~Liang, J.~Zhang, L.~Lu, and Z.~Ma, ``{Language-specific Boundary Learning for Improving Mandarin-English Code-switching Speech Recognition},'' in \emph{Proc. Interspeech 2023}, 2023, pp. 3322--3326.

\bibitem{audiosplicing2021}
C.~Du, H.~Li, Y.~Lu, L.~Wang, and Y.~Qian, ``Data augmentation for end-to-end code-switching speech recognition,'' in \emph{2021 IEEE Spoken Language Technology Workshop (SLT)}, 2021, pp. 194--200.

\bibitem{speechedit2023}
Z.~Liang, Z.~Song, Z.~Ma, C.~Du, K.~Yu, and X.~Chen, ``{Improving Code-Switching and Name Entity Recognition in ASR with Speech Editing based Data Augmentation},'' in \emph{Proc. Interspeech 2023}, 2023, pp. 919--923.

\bibitem{machineTranslate2021}
I.~Tarunesh, S.~Kumar, and P.~Jyothi, ``From machine translation to code-switching: Generating high-quality code-switched text,'' in \emph{Proceedings of the 59th Annual Meeting of the Association for Computational Linguistics and the 11th International Joint Conference on Natural Language Processing}.\hskip 1em plus 0.5em minus 0.4em\relax Association for Computational Linguistics, Aug. 2021, pp. 3154--3169.

\bibitem{yuhaibin2023}
H.~Yu, Y.~Hu, Y.~Qian, M.~Jin, L.~Liu, S.~Liu, Y.~Shi, Y.~Qian, E.~Lin, and M.~Zeng, ``Code-switching text generation and injection in mandarin-english asr,'' in \emph{ICASSP}, 2023, pp. 1--5.

\bibitem{LLM}
H.~Ke, S.~Tara, L.~Bo, and et~al., ``Improving multilingual and code-switching asr using large language model generated text,'' \emph{Automatic Speech Recognition and Understanding (ASRU)}, 2023.

\bibitem{selfsupervise_2023_1}
T.~Ogunremi, C.~Manning, and D.~Jurafsky, ``Multilingual self-supervised speech representations improve the speech recognition of low-resource {A}frican languages with codeswitching,'' in \emph{Proceedings of the 6th Workshop on Computational Approaches to Linguistic Code-Switching}.\hskip 1em plus 0.5em minus 0.4em\relax Association for Computational Linguistics, Dec. 2023, pp. 83--88.

\bibitem{selfsupervise_2023_2}
A.~Kulkarni, A.~Kulkarni, M.~Couceiro, and H.~Aldarmaki, ``{Adapting the adapters for code-switching in multilingual ASR},'' \emph{arXiv e-prints}, p. arXiv.2310.07423, Oct. 2023.

\bibitem{lae_2022}
J.~Tian, J.~Yu, C.~Zhang, Y.~Zou, and D.~Yu, ``{LAE: Language-Aware Encoder for Monolingual and Multilingual ASR},'' in \emph{Proc. Interspeech 2022}, 2022, pp. 3178--3182.

\bibitem{LAESTMoE}
G.~{Ma}, W.~{Wang}, Y.~{Li}, Y.~{Yang}, B.~{Du}, and H.~{Fu}, ``{LAE-ST-MoE: Boosted Language-Aware Encoder Using Speech Translation Auxiliary Task for E2E Code-switching ASR},'' \emph{Automatic Speech Recognition and Understanding (ASRU)}, 2023.

\bibitem{sw_wangyi_lr_moe_2023}
W.~Wang, G.~Ma, Y.~Li, and B.~Du, ``{Language-Routing Mixture of Experts for Multilingual and Code-Switching Speech Recognition},'' in \emph{Proc. Interspeech 2023}, 2023, pp. 1389--1393.

\bibitem{sw_pingan_2023}
F.~Tan, C.~Feng, T.~Wei, S.~Gong, J.~Leng, W.~Chu, J.~Ma, S.~Wang, and J.~Xiao, ``{Improving End-to-End Modeling For Mandarin-English Code-Switching Using Lightweight Switch-Routing Mixture-of-Experts},'' in \emph{Proc. Interspeech 2023}, 2023, pp. 4224--4228.

\bibitem{biencoder_luyizhou_2020}
Y.~Lu, M.~Huang, H.~Li, J.~Guo, and Y.~Qian, ``{Bi-Encoder Transformer Network for Mandarin-English Code-Switching Speech Recognition Using Mixture of Experts},'' in \emph{Proc. Interspeech 2020}, 2020, pp. 4766--4770.

\bibitem{SwitchTransformer}
W.~Fedus, B.~Zoph, and N.~Shazeer, ``Switch transformers: Scaling to trillion parameter models with simple and efficient sparsity,'' \emph{Journal of Machine Learning Research}, vol.~23, no. 120, pp. 1--39, 2022.

\bibitem{biencoder_2019}
S.~Zhang, Y.~Liu, M.~Lei, B.~Ma, and L.~Xie, ``{Towards Language-Universal Mandarin-English Speech Recognition},'' in \emph{Proc. Interspeech 2019}, 2019, pp. 2170--2174.

\bibitem{biencoder_2020}
X.~Zhou, E.~Yılmaz, Y.~Long, Y.~Li, and H.~Li, ``{Multi-Encoder-Decoder Transformer for Code-Switching Speech Recognition},'' in \emph{Proc. Interspeech 2020}, 2020, pp. 1042--1046.

\bibitem{biencoder_song22e_Interspeech}
T.~Song, Q.~Xu, M.~Ge, L.~Wang, H.~Shi, Y.~Lv, Y.~Lin, and J.~Dang, ``{Language-specific Characteristic Assistance for Code-switching Speech Recognition},'' in \emph{Proc. Interspeech 2022}, 2022, pp. 3924--3928.

\bibitem{sw_mole_2023}
Y.~Kwon and S.-W. Chung, ``Mole : Mixture of language experts for multi-lingual automatic speech recognition,'' in \emph{ICASSP}, 2023, pp. 1--5.

\bibitem{google_conformer_moe2023}
K.~Hu, B.~Li, T.~N. Sainath, Y.~Zhang, and F.~Beaufays, ``Mixture-of-expert conformer for streaming multilingual asr,'' \emph{ArXiv}, vol. abs/2305.15663, 2023.

\bibitem{U2++}
D.~Wu, B.~Zhang, C.~Yang, Z.~Peng, W.~Xia, X.~Chen, and X.~Lei, ``U2++: Unified two-pass bidirectional end-to-end model for speech recognition,'' \emph{ArXiv}, vol. abs/2106.05642, 2021.

\bibitem{ctc_attention}
S.~Watanabe, T.~Hori, S.~Kim, J.~R. Hershey, and T.~Hayashi, ``Hybrid ctc/attention architecture for end-to-end speech recognition,'' \emph{Neural Computation}, vol.~11, no.~8, p. 1240–1253, 2017.

\bibitem{conformer}
A.~Gulati, J.~Qin, C.-C. Chiu, N.~Parmar, Y.~Zhang, J.~Yu, W.~Han, S.~Wang, Z.~Zhang, Y.~Wu, and R.~Pang, ``{Conformer: Convolution-augmented Transformer for Speech Recognition},'' in \emph{Proc. Interspeech 2020}, 2020, pp. 5036--5040.

\bibitem{CTC}
A.~Graves, ``Connectionist temporal classification : labelling unsegmented sequence data with recurrent neural networks,'' \emph{Proc. Int. Conf. on Machine Learning, 2006}, 2006.

\bibitem{ASRU}
X.~Shi, Q.~Feng, and L.~Xie, ``The asru 2019 mandarin-english code-switching speech recognition challenge: Open datasets, tracks, methods and results,'' \emph{Proc. WSTCSMC}, 2020.

\bibitem{2015Librispeech}
V.~Panayotov, G.~Chen, D.~Povey, and S.~Khudanpur, ``Librispeech: An asr corpus based on public domain audio books,'' \emph{Proc. ICASSP}, 2015.

\bibitem{wenet}
Z.~Yao, D.~Wu, X.~Wang, B.~Zhang, F.~Yu, C.~Yang, Z.~Peng, X.~Chen, L.~Xie, and X.~Lei, ``Wenet: Production oriented streaming and non-streaming end-to-end speech recognition toolkit,'' in \emph{Proc. Interspeech}, 2021.

\bibitem{specaugment}
D.~S. Park, W.~Chan, Y.~Zhang, C.-C. Chiu, B.~Zoph, E.~D. Cubuk, and Q.~V. Le, ``{SpecAugment: A Simple Data Augmentation Method for Automatic Speech Recognition},'' in \emph{Proc. Interspeech 2019}, 2019, pp. 2613--2617.

\bibitem{wang21z_interspeech}
D.~Wang, S.~Ye, X.~Hu, S.~Li, and X.~Xu, ``{An End-to-End Dialect Identification System with Transfer Learning from a Multilingual Automatic Speech Recognition Model},'' in \emph{Proc. Interspeech 2021}, 2021, pp. 3266--3270.

\end{thebibliography}

\end{document}